\documentclass[sigconf,screen]{acmart}
\AtBeginDocument{%
  }

\usepackage{booktabs} 
\usepackage{array} 
\usepackage{makecell}
\usepackage{tabularx}
\usepackage{tcolorbox}
\usepackage{multirow}
\usepackage{pifont}
\usepackage{tikz}
\usepackage{ulem}
\usetikzlibrary{shapes.geometric, arrows, positioning, fit, calc, backgrounds, shadows}
\usepackage{subcaption}
\usepackage{algorithm}
\usepackage{algpseudocode}
\usepackage{float}
\usepackage[capitalize,noabbrev]{cleveref}
\usepackage{enumitem}

\setcopyright{cc}
\setcctype{by}
\acmDOI{10.1145/3832783.3837452}
\acmYear{2026}
\copyrightyear{2026}
\acmISBN{979-8-4007-2882-2/2026/10}
\acmConference[ASE '26]{Proceedings of the 41st IEEE/ACM International Conference on Automated Software Engineering}{October 12--16, 2026}{Munich, Germany}
\acmBooktitle{Proceedings of the 41st IEEE/ACM International Conference on Automated Software Engineering (ASE '26), October 12--16, 2026, Munich, Germany}
\acmSubmissionID{ase26main-p676-p}
\received{2026-03-26}
\received[accepted]{2026-06-18}

\begin{document}

\title{RepoProbe: Benchmarking Architecture-Aware Repository Comprehension with Checklists}

\author{%
  \mbox{Yuexi Yang\textsuperscript{2,*,\dag}} \quad
  \mbox{Alyssa Wu\textsuperscript{1,*,\ddag}} \quad
  \mbox{Ji Luo\textsuperscript{1}} \quad
  \mbox{Richeng Xuan\textsuperscript{1}} \\
  \mbox{Zhichao Hu\textsuperscript{1}} \quad
  \mbox{Yuhong Liu\textsuperscript{1}} \quad
  \mbox{Zhen Qin\textsuperscript{2,3,\S}}%
}
\affiliation{%
  \institution{%
    \mbox{\textsuperscript{1}Hunyuan, Tencent} \qquad
    \mbox{\textsuperscript{2}Zhejiang University} \qquad
    \mbox{\textsuperscript{3}Ningbo Global Innovation Center, Zhejiang University}%
  }
  \country{}
}
\email{yangyuexi@zju.edu.cn, alyssawwu@tencent.com, zhenqin@zju.edu.cn}


\begin{abstract}
The integration of Large Language Models (LLMs) into software engineering has shifted the focus from function-level generation to repository-scale assistance. However, existing benchmarks largely rely on bug reports from GitHub Issues, which often allow models to bypass genuine understanding via pattern matching on error logs. This misalignment under-measures \textbf{Edit Bias}, which refers to \textit{premature generation}, where models prematurely propose code modifications instead of understanding the existing repository architecture. Furthermore, current LLM-as-a-Judge scalar scoring suffers from high variance and low interpretability. This work introduces RepoProbe, a novel benchmark for evaluating repository-level code understanding through open-ended Q\&A using GitHub Discussions, which focuses on open-ended architectural inquiries rather than defect reporting. To ensure rigorous evaluation, we propose a Checklist-Based Verification Protocol that decomposes answers into atomic, verifiable facts, thereby replacing subjective ratings with objective verification. Our evaluation of state-of-the-art (SOTA) LLMs reveals a persistent gap between high clarity and evidence-grounded technical correctness. It also quantitatively confirms the prevalence of edit bias, in which models prioritize code generation instead of architectural analysis. Finally, we demonstrate that our verification protocol significantly improves evaluation reliability compared to traditional evaluations with scalar scoring.
\end{abstract}



\maketitle

\begingroup
\renewcommand{\thefootnote}{\fnsymbol{footnote}}
\footnotetext[1]{Both authors contributed equally to this research.}
\footnotetext[2]{Work done during an internship at Tencent.}
\footnotetext[3]{Project lead.}
\footnotetext[4]{Corresponding author.}
\endgroup

\section{Introduction}\label{sec:introduction}

AI-assisted software engineering has undergone a paradigm shift from function-level code completion to repository-scale autonomous software agents \citep{xia2025agentless,yang2024sweagent}. Modern Large Language Models (LLMs) are increasingly expected to implement features, refactor modules, and resolve issues in large codebases \citep{jimenez2024swebench}. In practice, this shifts the primary bottleneck from writing code to understanding repositories, tracing cross-file dependencies, and producing evidence-grounded explanations, making \textit{repository-level understanding} a critical reliability requirement \citep{bird2023taking}. Recent repository-scale studies suggest that reasoning over real codebases requires non-local integration across files and modules \citep{reporeason2026,fastcode2026}, while current evaluations of code agents remain primarily focused on bug fixing, under-examining repository explanation and realistic developer-facing understanding tasks \citep{beyondswe2026,locoeval2026}. 
Therefore, current evaluations provide only a partial view of this repository-level understanding capability. 
We highlight two limitations that become especially salient when the target skill is repository-level code understanding rather than localized patching.

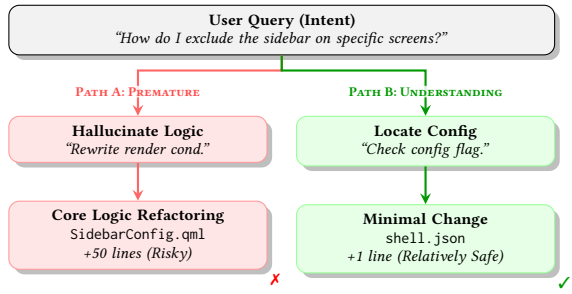
\begin{figure}[t]
\centering
\begin{tikzpicture}[
    node distance=0.6cm and 0.2cm, 
    font=\scriptsize, 
    box/.style={
        draw, rounded corners, align=center, 
        minimum height=0.6cm, 
        drop shadow, fill=white,
        inner sep=3pt 
    },
    input/.style={box, fill=gray!10, text width=7cm}, 
    bad/.style={box, fill=red!10, draw=red!40, text width=3.2cm},   
    good/.style={box, fill=green!10, draw=green!40, text width=3.2cm}, 
    arrow/.style={->, >=stealth, thick, color=gray},
    bad_arrow/.style={->, >=stealth, thick, color=red!60},
    good_arrow/.style={->, >=stealth, thick, color=green!60!black},
    label_node/.style={font=\bfseries\tiny}, 
    label_bg/.style={label_node, fill=white, fill opacity=0.85, text opacity=1, rounded corners=1pt, inner sep=0.5pt}
]

\node[input] (intent) {
    \textbf{User Query (Intent)} \\
    \textit{``How do I exclude the sidebar on specific screens?''}
};

\node[bad, below=0.8cm of intent, xshift=-1.9cm] (process_bad) {
    \textbf{Hallucinate Logic} \\
    \textit{``Rewrite render cond.''}
};
\node[coordinate, above=0.25cm of process_bad, yshift=2pt] (label_bad) {};

\node[bad, below=0.5cm of process_bad] (result_bad) {
    \textbf{Core Logic Refactoring} \\
    \texttt{SidebarConfig.qml} \\
    \textit{+50 lines (Risky)}
};

\node[good, below=0.8cm of intent, xshift=1.9cm] (process_good) {
    \textbf{Locate Config} \\
    \textit{``Check config flag.''}
};
\node[coordinate, above=0.25cm of process_good, yshift=2pt] (label_good) {};

\node[good, below=0.5cm of process_good] (result_good) {
    \textbf{Minimal Change} \\
    \texttt{shell.json} \\
    \textit{+1 line (Relatively Safe)}
};

\draw[bad_arrow] (intent.south) -- ++(0,-0.2) -| (process_bad.north);
\draw[good_arrow] (intent.south) -- ++(0,-0.2) -| (process_good.north);
\draw[bad_arrow] (process_bad) -- (result_bad);
\draw[good_arrow] (process_good) -- (result_good);

\node[label_bg, text=red!70] at (label_bad) {\textsc{Path A: Premature}};
\node[label_bg, text=green!60!black] at (label_good) {\textsc{Path B: Understanding}};

\node[anchor=north west, xshift=-2pt, yshift=2pt] at (result_bad.south east) {\color{red}\ding{55}};
\node[anchor=north west, xshift=-2pt, yshift=2pt] at (result_good.south east) {\color{green!60!black}\ding{51}};

\end{tikzpicture}
\caption{Without appropriate repository understanding, models exhibit \textit{premature generation} (Left) by refactoring source code for a requirement that only needs simple configuration changes (Right). This \textit{implementation--configuration mismatch} incurs unnecessary engineering cost.}
\label{fig:qml_example}
\end{figure}

\textbf{Limitation 1: Existing benchmarks confound understanding with defect localization.} 
Figure~\ref{fig:qml_example} illustrates a critical failure mode often masked by current evaluation paradigms: a scenario where a user request requires only a minimal configuration change, yet the model---lacking a holistic mental model of the repository---proposes an unnecessary and high-risk refactoring of core logic. 
We term this phenomenon \textbf{Edit Bias}, referring to premature generation, which typically emerges when models address open-ended inquiries without explicit localization cues.
This example highlights the limitations of the dominant defect-centric benchmarks such as SWE-bench \citep{jimenez2024swebench}. 
Such benchmarks typically provide strong localization cues (e.g., stack traces, error logs, and filenames), enabling a debugging shortcut. 
Models can exploit these cues to match surface patterns and navigate directly to suspicious files, bypassing the need to understand the repository itself. 
Consequently, while models may achieve high pass rates on bug-fixing tasks, cases like Figure~\ref{fig:qml_example} reveal that they often struggle to answer open-ended questions about repository behavior, design rationale, and inter-module dependencies when those shortcuts are removed.

\textbf{Limitation 2: Prevailing evaluation protocols are ill-suited for open-ended repository explanations.} When the task is reframed as ``explain how the repository works'' rather than ``make tests pass'', evaluation becomes an oracle problem: there is no single executable signal, and correctness depends on multiple distributed facts. The prevailing LLM-as-a-Judge approach using holistic scalar scores \citep{zheng2024judging} is appealing for its simplicity, but it suffers from practical limitations that are amplified in repository-level settings. First, scalar scores are often opaque and provide weak diagnostic value for model improvement. Second, they can be unstable across repeated runs, and are susceptible to known biases such as favoring longer, more fluent answers regardless of technical correctness \citep{singhal2023long}. Recent benchmark efforts have also begun to explore more structured evaluation for coding-assistant question answering \citep{rubberduckbench2026}. These limitations become more consequential precisely when we move away from localized bug repair toward open-ended repository understanding, where partial correctness and evidence grounding are paramount.

\textbf{To overcome these limitations, we introduce RepoProbe.} RepoProbe is a \textbf{discussion-based repository understanding benchmark} grounded in real GitHub Discussions. It addresses the data gap by shifting from traditional GitHub Issues-based datasets to naturally occurring ``how'' and ``why'' inquiries that require semantic mapping from user intent to repository structure \citep{hata2022github}. To simultaneously improve evaluation stability and interpretability, we implement a \textbf{Checklist-Based Verification Protocol}. Instead of relying on opaque scalar scores, this protocol maps open-ended repository answers to discrete, verifiable technical facts, enabling more rigorous and reproducible measurement.

RepoProbe complements, rather than replaces, recent repository-level benchmarks and methods. In particular, RepoReason \citep{reporeason2026} evaluates repository-level reasoning through abductive assertion verification, asking models to reconstruct program states from masked unit-test assertions. While this setting is valuable for studying execution reasoning, recovering an assertion value does not show that a model can answer the kinds of questions that developers actually ask in practice, such as how a repository is organized, why a component behaves in a certain way, or what part of the system is relevant to a user-facing need. RepoProbe uses a complementary Q\&A task formulation to evaluate repository understanding in real-world discussions, making it closer to realistic developer-facing usage scenarios.

This work makes the following contributions:

\begin{itemize}[leftmargin=2em]
    \item \textbf{Discussion-based repository understanding benchmark.} We introduce RepoProbe, a benchmark that evaluates repository-level code understanding through open-ended question-answer\-ing task from real GitHub Discussions, targeting open-ended repository questions, requiring cross-file evidence gathering beyond defect localization shortcuts.

    \item \textbf{Checklist-Based Verification Protocol.} We propose a check\-list-based verification protocol that decomposes open-ended answers into weighted, verifiable technical items with fine-grained rubrics. Experiments on our benchmark show that this protocol improves evaluation reliability compared to traditional scalar scoring.

    \item \textbf{Empirical Characterization of Edit Bias.} Through quantitative and qualitative analyses, we show that frontier models can exhibit edit bias in open-ended repository inquiries, where fluent explanations mask a lack of evidence-grounded correctness, This bias is shown to account for a substantial fraction of failure cases.
\end{itemize}

\section{Related Work}\label{sec:related_work}

\subsection{Repository-Level Benchmarks for Code Understanding}

Recent evaluation efforts have moved beyond function-level code completion toward repository-scale settings, where models must reason over long contexts, cross-file dependencies, and project structure. One line of work emphasizes \textbf{long-context code understanding}. Benchmarks such as RepoQA \citep{liu2024repoqa}, LongCodeU \citep{li2025longcodeu}, and LoCoBench \citep{locobench2025} stress whether models can locate relevant code, maintain performance over long repositories, and reason under large context windows. These benchmarks provide important evidence that long-context access remains a bottleneck. However, their primary focus is still on retrieval, localization, or long-context processing, rather than on answering realistic developer questions about repository behavior and design.

\begin{table*}[t!]
    \centering
    \caption{Comparison Between RepoProbe and Representative Code Understanding Benchmarks}
    \label{tab:benchmark_comparison}
    \footnotesize
    \setlength{\tabcolsep}{4pt}
    \renewcommand{\arraystretch}{1.08}
    \begin{tabular}{>{\raggedright\arraybackslash}p{2.2cm}>{\raggedright\arraybackslash}p{2.8cm}>{\raggedright\arraybackslash}p{3.0cm}cccc}\toprule
         \makecell[l]{Benchmark}& \makecell[l]{Source}& \makecell[l]{Primary Evaluation\\Target}& \makecell[c]{Repo-\\Level?}& \makecell[c]{Multi-\\lingual?} &\makecell[c]{Open-ended\\Understanding?}&\makecell[c]{Checklist-Based\\Verification?}\\\midrule
         CodeXGlue \cite{lu2024codexglue}&  Existing datasets&  General Code Understanding& \ding{55} &  $\checkmark$ &\ding{55} &\ding{55} \\
         xCodeEval \cite{khan2024xcodeeval}&  CodeForces&  General Code Understanding& \ding{55} &  $\checkmark$ &\ding{55} &\ding{55} \\
         CRUXEval \cite{gu2024cruxeval}& LLM-generated& Execution Reasoning& \ding{55}& \ding{55} &\ding{55} &\ding{55} \\
         RepoQA \cite{liu2024repoqa}&  GitHub repositories& Long-Context Retrieval& $\checkmark$& $\checkmark$ &\ding{55} &\ding{55} \\
         LongCodeU \cite{li2025longcodeu}& GitHub repositories& Long-Context Retrieval& $\checkmark$& \ding{55} &\ding{55} &\ding{55} \\
         LoCoBench \cite{locobench2025}& LLM-generated& General Code Understanding& $\checkmark$& $\checkmark$ &\ding{55} &\ding{55} \\
         CoReQA \cite{coreqa2025}& GitHub Issues and Comments& Repository QA& $\checkmark$& $\checkmark$ &$\checkmark$ &\ding{55} \\
         SWE-QA \cite{peng2025sweqa}& GitHub repositories& Repository QA& $\checkmark$& \ding{55} &$\checkmark$ &\ding{55} \\
         RubberDuckBench \cite{rubberduckbench2026}& GitHub pull request comments& Coding Assistant QA& \ding{55} & $\checkmark$ &$\checkmark$ &$\checkmark$ \\
         RepoReason \cite{reporeason2026}& GitHub repositories + unit tests& Assertion Verification& $\checkmark$& \ding{55} &\ding{55} &\ding{55} \\
         LoCoEval \cite{locoeval2026}& LLM-generated conversations& Conversation Management& $\checkmark$& \ding{55} &$\checkmark$ &\ding{55} \\
         BeyondSWE \cite{beyondswe2026}& Real-world software tasks& Issue Resolution& $\checkmark$& \ding{55} &\ding{55} &\ding{55} \\
         \textbf{RepoProbe (Ours)}& GitHub Discussions& Repository QA& $\checkmark$& $\checkmark$ &$\checkmark$ &$\checkmark$ \\ \bottomrule
    \end{tabular}
    \vspace{-4pt}
\end{table*}

Another major line of work evaluates \textbf{repository-level software engineering tasks} such as bug fixing, repair, and refactoring. SWE-bench \citep{jimenez2024swebench}, BeyondSWE \citep{beyondswe2026}, and SWE-Refactor \citep{swerefactor2026} are representative examples. These benchmarks offer strong ecological validity because they are grounded in real repositories and practical engineering tasks. At the same time, they often provide explicit task-completion cues, including issue descriptions, failing behaviors, or concrete edit goals. As a result, strong performance in these settings does not necessarily imply that a model can explain repository organization, recover design rationale, or answer open-ended questions when such localization cues are absent. Prior repair-oriented work also suggests that broader repository context matters substantially for solving real code tasks \citep{ehsani2025hierarchical}, further underscoring the difference between localized patching and genuine repository comprehension.

Closer to our target setting are recent \textbf{repository-level question answering and reasoning} benchmarks. CoReQA \citep{coreqa2025} and SWE-QA \citep{peng2025sweqa} explicitly study repository-level question answering, while RepoReason \citep{reporeason2026} evaluates agentic repository reasoning through abductive assertion verification. These benchmarks show that repository understanding requires more than local search: models must aggregate evidence across files, infer state, and reason about non-local interactions. Nevertheless, their task formulations differ from the setting targeted by RepoProbe. In particular, RepoReason measures execution-oriented state reconstruction, whereas RepoProbe evaluates whether a model can answer naturally occurring repository questions grounded in developer discussions. Therefore, our benchmark complements these efforts by focusing on open-ended ``how'' and ``why'' questions derived from GitHub Discussions, where the central challenge is semantic mapping from user intent to repository evidence rather than assertion recovery or issue resolution. To provide a clearer overview, \Cref{tab:benchmark_comparison} summarizes representative benchmarks and contrasts them along key dimensions relevant to repository-level understanding.

\subsection{Repository-Level Reasoning and Navigation}

Beyond benchmark construction, several recent studies frame repo\-sitory-scale performance as a problem of \textbf{navigation, exploration, and evidence aggregation}. FastCode \citep{fastcode2026} argues that effective repository understanding requires structural scouting and cost-aware context construction instead of naively consuming raw context. SWE-Adept \citep{sweadept2026} likewise highlights the importance of codebase analysis through tool-augmented agentic workflows. These works suggest that repository understanding is not merely a matter of fitting more tokens into context, but of selecting and integrating the right evidence from a large codebase.

Other work focuses on \textbf{reasoning abstractions} for repository-scale software engineering. Agentic Code Reasoning \citep{agenticcodereasoning2026} studies semi-formal reasoning for code analysis, while RepoRepair \citep{reporepair2026} leverages repository documentation to support repair in large codebases. Together with RepoReason \citep{reporeason2026}, these studies reinforce a common conclusion: repository-level performance depends on cross-file integration, state tracking, and structured evidence use rather than simple retrieval alone. RepoProbe is designed to evaluate this capability under a developer-facing QA setting, where agents must navigate the repository and synthesize an evidence-grounded natural-language answer instead of merely returning a location or producing a patch.

\subsection{Evaluation of Open-Ended Code QA}
\label{sec:evaluation_metrics}

Evaluating open-ended answers in the code domain remains difficult. Classical lexical-overlap metrics such as BLEU and ROUGE are known to be poorly aligned with technical correctness because correct explanations may differ substantially from a reference in wording while preserving the same facts \citep{chen2021evaluating}. For this reason, recent work increasingly adopts \textbf{LLM-as-a-Judge} scoring, as popularized in broader language-model evaluation \citep{zheng2024judging}. However, scalar judging is vulnerable to instability and known biases, including a tendency to reward verbosity and surface fluency \citep{singhal2023long}, which becomes especially problematic for repository-level questions whose answers are multi-fact, partial-credit, and evidence-dependent.

To address these limitations, recent evaluation work has begun to explore \textbf{structured verification}. CheckEval \citep{lee2025checkeval} shows the value of decomposing open-ended answers into checklist-style criteria, and RubberDuckBench \citep{rubberduckbench2026} brings rubric-based evaluation into the code-assistant setting using real-world coding questions. This line of work is particularly relevant to repository-level QA, where the oracle is inherently compositional: an answer may be partially correct, omit important technical facts, or provide a fluent but unsupported explanation. RepoProbe builds on this emerging direction by introducing a checklist-based verification protocol tailored to repository understanding tasks, enabling finer-grained and more reproducible evaluation than holistic scalar scores.

\smallskip
Overall, prior work establishes the importance of long-context processing, repository navigation, and structured evaluation, but leaves a gap at their intersection. Existing benchmarks either emphasize retrieval, issue resolution, or execution-oriented reasoning, while open-ended repository QA remains comparatively underexplored and difficult to evaluate reliably. RepoProbe addresses this gap by combining discussion-derived repository questions with checklist-based verification for evidence-grounded, open-ended repository understanding.

\section{RepoProbe Benchmark}

\subsection{Task Definition}\label{sec:task_definition}

RepoProbe evaluates repository-level code understanding through \textbf{repository-level Q\&A}: a developer-facing comprehension task in which an agent answers questions about an existing codebase rather than modifying it. Formally, the task takes as input a target repository $R$ and a natural-language question $q$, and requires the model to produce a textual answer $a = f(R, q)$. The answer should resolve the developer's information need by explaining repository behavior, design rationale, API usage, configuration choices, or relevant implementation details.

The evaluation is separate from the task itself. Given a generated answer $a$ and reference material $M$, an evaluation mechanism $\mathrm{Score}(a, M)$ assigns a score indicating how well the answer covers the required technical facts. In RepoProbe, $q$ is a self-contained question derived from a real GitHub Discussion, and the reference material $M$ consists of the developer-accepted reference answer $a_{\text{ref}}$ and a weighted checklist $\mathcal{C}$ derived from $q$ and $a_{\text{ref}}$. Each instance follows a fixed four-part structure---question, repository, developer-accepted reference answer, and weighted checklist---so that the input, output, and assessment target are clearly bounded.

This setup evaluates whether the answer demonstrates semantic mapping from user intent to repository evidence: navigating the repository, identifying relevant files or components, tracing cross-file relationships, reasoning over long-range code context, and synthesizing an explanation grounded in the repository. These capabilities distinguish repository-level Q\&A from program repair and refactoring tasks \citep{jimenez2024swebench,swerefactor2026}, where the model is expected to change the codebase and success is judged by executable outcomes such as passing tests or satisfying an edit request. In contrast, repository-level Q\&A requires the model to explain the existing system rather than modify it \citep{coreqa2025,peng2025sweqa}.



\subsection{Data Curation} \label{sec:data_curation}

\begin{figure*}
    \centering
    \includegraphics[width=1\linewidth]{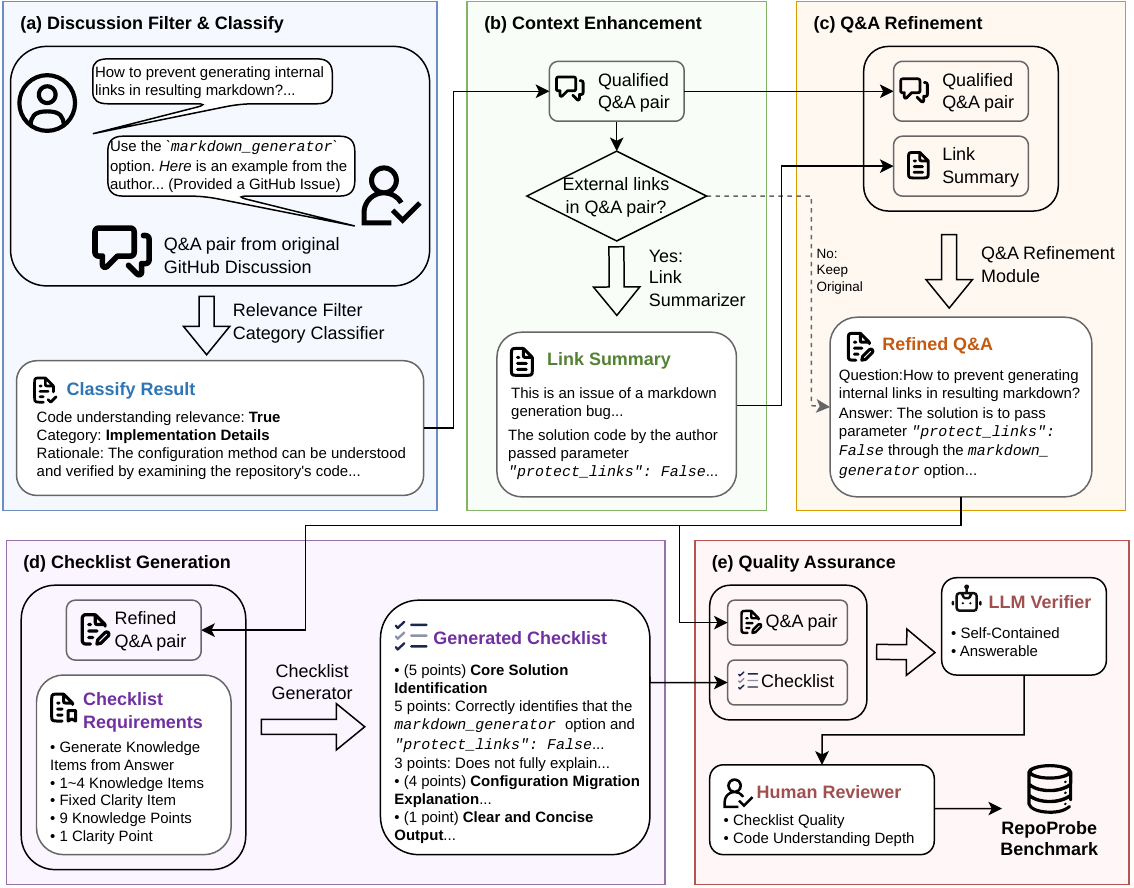}
    \caption{RepoProbe data curation pipeline from answered GitHub Discussions to self-contained, checklist-scored Q\&A tasks. Hollow arrows represent LLM-driven steps.}
    \label{fig:construction_pipeline}
    \vspace{-4pt}
\end{figure*}

To construct a benchmark that emphasizes deep code understanding, we source our data from GitHub Discussions rather than traditional GitHub Issues. Issues often contain explicit localization cues such as stack traces that bypass the need for repository-level understanding. In contrast, Discussions typically feature abstract architectural goals without referencing specific files, requiring agents to perform \textit{Semantic Mapping} via cross-file exploration. To transform these raw, noisy community discussions into structured, self-contained, and objectively scorable Q\&A pairs, we designed the multi-stage curation pipeline illustrated in Figure~\ref{fig:construction_pipeline}.

\textbf{Repository Selection and Raw Data Harvesting.} To capture recent development trends and reduce the risk of training-data contamination, we imposed strict criteria on the source repositories. We selected GitHub projects that (1) were created on or after January 1, 2024, (2) have accumulated at least 1,000 stars, indicating substantial community adoption, and (3) use the \textit{Discussions} feature for developer support, with “Q\&A” or “Help” categories and at least ten discussions explicitly marked as “answered” by maintainers or original authors. Applying these criteria yielded a set of popular and actively maintained projects. From these repositories, we collected all answered discussions, resulting in an initial corpus of 5,233 raw discussion threads.

\textbf{Data Cleaning and Relevance Filtering.} Raw GitHub discussions contain a wide variety of topics, many of which are not directly related to code understanding, such as release announcements, or purely conceptual debates. We first removed discussions where the question or the accepted answer relied on images, since our benchmark targets text-based language model evaluation. We then used \texttt{Claude Sonnet 4.5} to classify the remaining discussions according to their relevance to code understanding. We defined three code-understanding categories grounded in established program-comprehension knowledge types \citep{maalej2014comprehension,fritz2012developers}: (i) \textit{Project Architecture}, focusing on repository structure and inter-module relationships; (ii) \textit{Business Logic}, focusing on functional behavior and data flow; and (iii) \textit{Implementation Details}, focusing on concrete APIs, code snippets, or configuration details. This relevance filtering reduced the corpus to 2,613 discussions with clear code-understanding intent.

\textbf{Context Enhancement and Q\&A Refinement.} A major challenge in mining developer discussions is that questions often rely on external context, such as links to code snippets, documentation, or other issues that are not immediately visible in the text. To address this, we implemented a context enhancement workflow:

\begin{figure*}[t]
    \centering
    \begin{minipage}{0.32\linewidth}
        \centering
        \includegraphics[width=\linewidth]{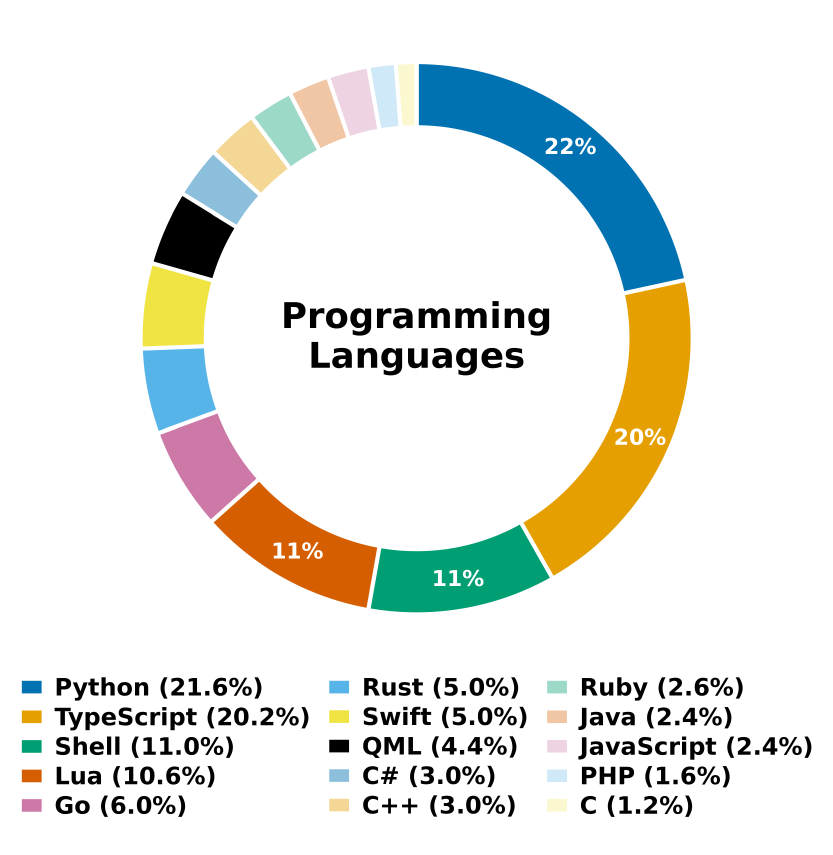}
    \end{minipage}
    \hfill
    \begin{minipage}{0.32\linewidth}
        \centering
        \includegraphics[width=\linewidth]{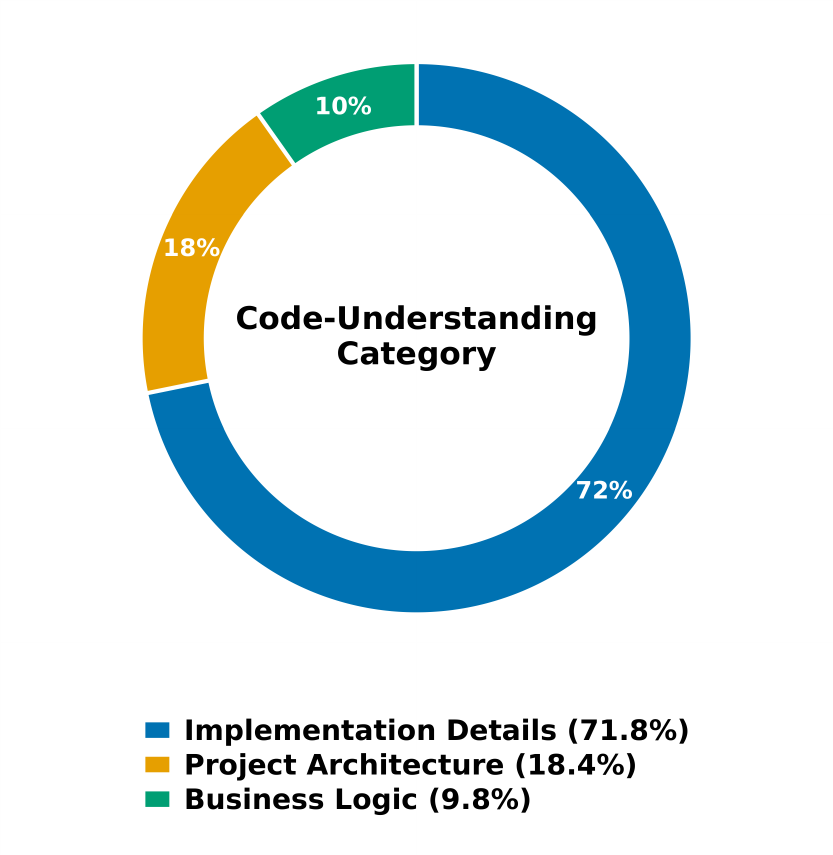}
    \end{minipage}
    \hfill
    \begin{minipage}{0.32\linewidth}
        \centering
        \includegraphics[width=\linewidth]{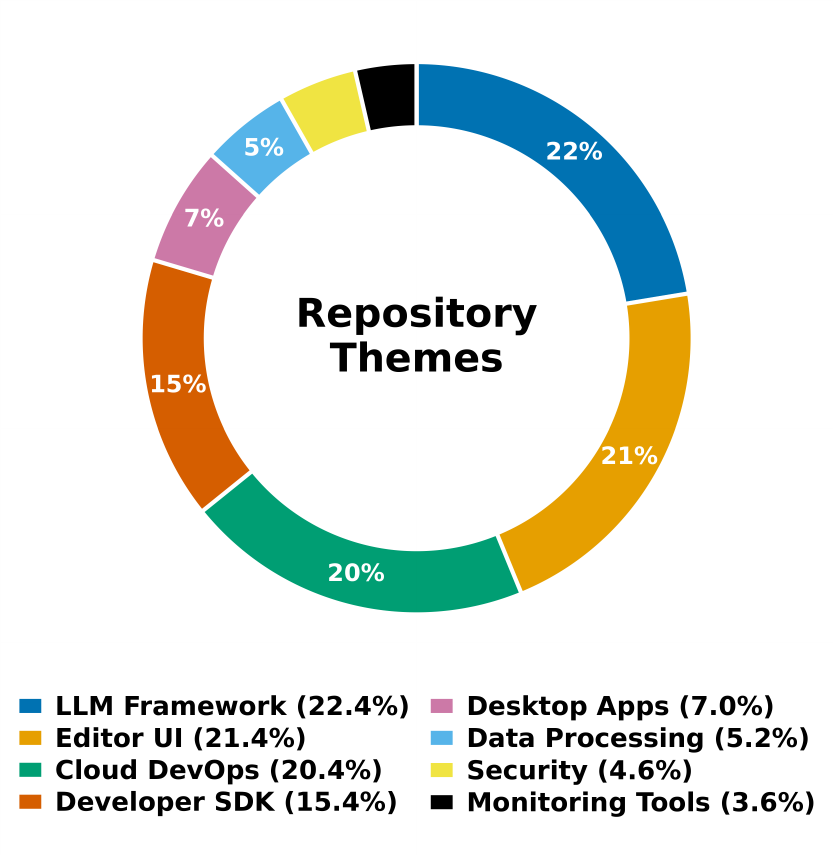}
    \end{minipage}
    \vspace{-3pt}
    \caption{Distribution of RepoProbe questions by programming language (left), code-understanding category (center), and repository theme (right).}
    \label{fig:question_distribution_overview}
\end{figure*}

\begin{enumerate}
\item \textit{Extraction:} We extracted the title and body of the discussion as the Question and the accepted answer as the Answer.
\item \textit{Context Resolution:} We identified external URLs within the text. We excluded internal references such as \textit{localhost} and crawled valid external links to retrieve their content. \texttt{Claude Sonnet 4.5} was then used to verify the relevance of the crawled content to the repository and summarize it.
\item \textit{Content Integration:} For Q\&A pairs containing external links, we used \texttt{Claude Sonnet 4.5} to integrate the summarized external content directly into the Question and Answer text. The model was instructed to preserve the original information while incorporating the external context, ensuring the Q\&A becomes \textbf{self-contained and solvable without internet access}. Q\&A pairs without external links were retained in their original form to preserve authentic developer communication.
\end{enumerate}

This stage transforms context-fragmented discussions (those with external dependencies) into standalone Q\&A pairs that can be presented directly to models. Crucially, after this integration step, agents evaluating on RepoProbe only need access to the repository code and the processed question text—no external network access is required during evaluation.

\textbf{Evaluation Checklist Generation.} To enable objective and reproducible evaluation under a unified scoring interface, \texttt{Claude Sonnet 4.5} analyzes the refined Q\&A pair with a checklist-generation prompt and output schema. The prompt requires the model to produce reference-derived knowledge items, integer score weights, and discrete rubrics under the standardized 10-point structure. Checklist item content therefore varies with each question and reference answer, but the generation pipeline itself remains the same for every instance. This mechanism converts open-ended answers into machine-checkable scoring procedures. We detail the checklist structure and our verification protocol in Section~\ref{sec:metrics}.

\textbf{Quality Assurance.} We next applied automatic and human-in-the-loop quality control to remove problematic or ambiguous cases. Using a verification protocol based on \texttt{Claude Opus 4.5}, we performed two critical checks: (1) \textit{External Link Validation}, discarding Q\&A pairs whose correctness depended on external links that could not be reliably resolved or whose relevance to the repository could not be verified; and (2) \textit{Answerability Check}, verifying that the question is answerable and that the evidence required by each checklist item can be obtained from only the repository's code and the information explicitly present in the rewritten question. After this verification step, we obtain 1,107 candidate Q\&A pairs from 51 repositories.

\begin{table}[t]
\centering
\caption{RepoProbe Benchmark Statistics}
\label{tab:benchmark_statistics}
\begin{tabular}{lc}
\toprule
\textbf{Statistic} & \textbf{Value} \\
\midrule
Total Repositories & 50 \\
Total Q\&A Pairs & 500 \\
Programming Languages & 15 \\
Avg. Files per Repo & $\sim$1,200 \\
Avg. LoC per Repo & $\sim$380k \\
\bottomrule
\end{tabular}
\vspace{-5pt}
\end{table}
 
A subset of the remaining samples was then manually inspected by expert annotators. They applied four orthogonal validation criteria: \textit{QA consistency} (the accepted answer must directly and non-contradictorily answer the question), \textit{checklist quality} (items must be correct, sufficiently granular, and non-conflated), \textit{task validity} (the question must require repository context rather than generic advice), and \textit{code relevance} (the answer must depend on code or configuration facts). Annotator disagreements were resolved through discussion before finalizing each sample. More than half of the candidate samples were discarded during this stage, with the dominant cause being direct answerability without genuine repository understanding. After this stage, we obtained a pool of 500 validated code-understanding Q\&A pairs drawn from 50 active repositories.


\subsection{Statistics of RepoProbe benchmark}

\Cref{tab:benchmark_statistics} summarizes the main statistics of RepoProbe. Each sample in the benchmark is a real GitHub Discussion question rewritten into a self-contained Q\&A instance with an executable checklist. Among the final 500 samples, 225 include external context that was merged into the question or answer text to support offline evaluation. The underlying repositories are large, multi-module projects rather than isolated code snippets: on average, each repository contains $\sim$1,200 files and $\sim$380k lines of code, making many tasks inherently long-context and dependent on cross-file evidence gathering and dependency tracing.

\Cref{fig:question_distribution_overview} further characterizes the benchmark along three axes that later motivate our stratified evaluation in Section~\ref{sec:evaluation}. RepoProbe spans 15 programming languages with a long-tailed distribution, where a few languages account for a large fraction of questions, followed by a diverse set of lower-frequency ecosystems. Also, the benchmark covers multiple code-understanding categories and repository themes, enabling the analysis of model performance from complementary structural and domain-specific perspectives.

\section{Checklist-Based Verification Protocol} \label{sec:metrics}

As detailed in Section~\ref{sec:evaluation_metrics}, traditional scalar scoring acts as a ``fuzzy oracle'' with high variance and low interpretability. To address these limitations, we introduce a \textbf{Checklist-Based Verification Protocol} as a core contribution. This approach decomposes open-ended repository evaluation into atomic, verifiable technical checks.

\subsection{Checklist Structure}\label{sec:structure}
Following the data curation methodology (Section~\ref{sec:data_curation}), each question $q$ is accompanied by a reference answer and a checklist $\mathcal{C} = \{c_1, \ldots, c_n\}$. Each item $c_i$ specifies three components:
\begin{enumerate}[leftmargin=2em]
  \item \textbf{Key Point}: A precise, atomic aspect of the reference answer, such as \textit{identification of root cause} or \textit{API usage correctness}.
  \item \textbf{Score Weight}: The maximum points allocated to this aspect.
  \item \textbf{Evaluating Rubric}: A tiered criterion mapping partial correctness to discrete scores.
\end{enumerate}

All checklists are generated with a unified prompt template and standardized schema across the entire benchmark. The schema fixes the required fields for every item including key point, score weight, and discrete evaluating rubric as well as the total score budget and judge output format. The checklist content therefore varies with the specific question and reference answer, but the generation procedure, score scale, item fields, valid-score constraints, and verification interface are shared by all instances.

We adopt a standardized 10-point scale. Each \textbf{RepoProbe} question has $k \in \{1, 2, 3, 4\}$ content-specific knowledge items derived from the reference answer, with integer weights summing to 9. The remaining 1 point is reserved for a universal \textbf{Clarity \& Organization} criterion that flags severe hallucinations or incoherent structures. For every item, the valid score values are fixed during dataset construction and released as part of the checklist; the evaluator may only choose from these predefined values.

For knowledge items, we use two types of rubrics: \textbf{Multi-tier rubrics} for complex items (e.g., 4 points for fully correct, 2 for partially correct, 0 for incorrect) and \textbf{Binary rubrics} for single-fact items (full credit or zero). The total score for a generated response is simply the sum of the discrete scores assigned to each checklist item by the LLM judge. In our evaluation, we report this as a normalized score rate (\%).

\subsection{Constrained LLM Evaluation via Rationale-Guided Scoring}
We formulate the evaluation task as a checklist-based verification problem. In contrast to scalar scoring, which relies on a single opaque inference step, our protocol decomposes the judgment into a sequence of item-wise decisions.

To improve the transparency and stability of the judgments, our verification protocol enforces two constraints on the LLM evaluator:
\begin{enumerate}[leftmargin=2em]
  \item \textbf{Rationale-first constraint:} The LLM is instructed to first generate a textual rationale (serving as evidence) and then decide the score based on this rationale. This leverages Chain-of-Thought reasoning to ground decisions in specific details of the generated response, reducing post-hoc rationalization.
  \item \textbf{Rubric adherence:} The assigned score must be strictly chosen from the valid discrete set defined by the rubric. This prevents arbitrary scalar assignments and enforces consistency across repeated evaluations.
\end{enumerate}

\textbf{Implementation Strategy.} For efficiency, we implement this verification as a single-pass LLM inference. The checklist is serialized into a structured schema, and the LLM outputs a strictly formatted JSON sequence of rationales and scores for all items simultaneously. This design trades strict step-wise independence for efficiency while allowing the model to use global context from the full answer. In our experiments, we demonstrate that this constrained verification significantly reduces variance and provides more consistent rationales than scalar scoring baselines.
\section{Experimental Evaluation} \label{sec:evaluation}
To rigorously evaluate repository-level understanding, we move beyond static prompting and adopt an agentic evaluation paradigm. This section describes our experimental setup and reports results for our research questions. Unless otherwise stated, we keep the agentic environment fixed across models to isolate differences in model capability rather than differences in scaffolding.

Building on the evaluation paradigm above, our experiments are designed to answer the following research questions, each reported in a dedicated subsection: 
\begin{itemize}[leftmargin=2em]
\item \textbf{RQ1} (Section~\ref{sec:rq1}): How well do frontier LLMs demonstrate repository-level code understanding on Q\&A tasks, and how does performance vary across programming languages, code-understanding categories, and repository themes?
\item \textbf{RQ2} (Section~\ref{sec:rq2}): Does checklist-based verification provide more stable and interpretable evaluation signals than traditional LLM-as-a-Judge scalar scoring for repository-level Q\&A tasks?
\item \textbf{RQ3} (Section~\ref{sec:rq3}): To what extent do frontier LLMs exhibit \textit{Edit Bias} when answering discussion-based repository-level architectural inquiries, and what qualitative patterns characterize this failure mode?
\end{itemize}

\subsection{Experimental Setup} \label{sec:experiment_setup}
\subsubsection{Evaluation Framework: Agentic Scaffolding}
Unlike traditional code benchmarks that provide models with a truncated context window, RepoProbe requires navigating vast, multi-file codebases. To simulate a real-world developer environment, we evaluate all models under a fixed agentic scaffolding \textbf{Claude Code}, which provides tools for file system navigation, symbol searching, and cross-reference tracing.


\subsubsection{The Selection of LLMs} \label{sec:model_selection}

To ensure that our evaluation reflects the capabilities encountered in realistic software-engineering workflows, we select models according to three criteria. First, we include \textbf{frontier models from multiple major providers} rather than focusing on a single ecosystem, so that the benchmark reflects the competitive landscape actually faced by developers. Second, we include models representing \textbf{two different SOTA time points}: an earlier frontier generation and a more recent frontier generation. This design allows us to assess not only absolute capability, but also whether progress across release cycles consistently translates into better repository-level understanding. We divide the evaluated systems into two broad groups:

\paragraph{Closed-weight Models.} This cohort covers proprietary frontier systems from multiple providers. It includes models from \textbf{OpenAI} (\texttt{GPT-5.2}~\cite{openai2025gpt52}, \texttt{GPT-5.4}~\cite{openai2026gpt54}), \textbf{Anthropic} (\texttt{Claude Opus 4.5}~\cite{anthropic2025opus45}, \texttt{Claude Opus 4.6}~\cite{anthropic2026opus46}, \texttt{Claude Sonnet 4.6}~\cite{anthropic2026sonnet46}), \textbf{Google} (\texttt{Gemini 3 Flash}~\cite{deepmind2025gemini3flash}, \texttt{Gemini 3 Pro}~\cite{deepmind2025gemini3pro}, \texttt{Gemini 3.1 Pro}~\cite{deepmind2026gemini31pro}), \textbf{xAI} (\texttt{Grok 4.20}~\cite{xai2026grok420}), \textbf{Qwen} (\texttt{Qwen3-Max}~\cite{qwen2025qwen3max}, \texttt{Qwen3.5-Plus}~\cite{qwen2026qwen35}), and \textbf{ByteDance} (\texttt{Doubao-Seed-1.8}~\cite{bytedance2025seed18}, \texttt{Doubao-Seed-2.0}~\cite{bytedance2026seed20}). 

\paragraph{Open-weight Models.} This cohort covers leading open-weight or openly available models that are widely used in practice. It includes models from \textbf{DeepSeek} (\texttt{DeepSeek-V3.2}~\cite{deepseek2025v32}), \textbf{MiniMax} (\texttt{MiniMax-M2.1}~\cite{minimax2025m21}, \texttt{MiniMax-M2.5}~\cite{minimax2026m25}), \textbf{Zhipu AI} (\texttt{GLM-4.7}~\cite{glm2025glm47}, \texttt{GLM-5}~\cite{zhipu2026glm5}), and \textbf{Moonshot AI} (\texttt{Kimi-K2}~\cite{kimiteam2025kimik2}, \texttt{Kimi-K2.5}~\cite{kimiteam2026kimik25}).

Overall, this selection spans 20 models from 10 providers and provides a broad snapshot of the current landscape of LLM-based software engineering agents. Many of these models are also surfaced in practical coding environments, making the evaluation directly relevant to real developer workflows.

\subsubsection{Evaluation Metrics}
 
 All model answers are graded using our \textbf{Checklist-Based Verification Protocol} (Section~\ref{sec:metrics}). This design makes evaluation verifiable at the level of atomic technical items, and supports fine-grained diagnosis beyond scalar scoring.

 Unless otherwise stated, we use \texttt{Claude Sonnet 4.5} as the judge model for checklist verification and keep it fixed across all evaluated systems and repeated runs to control for judge variability. Because \texttt{Claude Sonnet 4.5} is also used for checklist generation during dataset construction, we deliberately exclude it from the evaluated model set to avoid self-evaluation.

We report four summary metrics, computed as averages over either the full benchmark or a stratified subset of questions. Let $s \in [0,10]$, $s_{\text{kn}} \in [0,9]$, and $s_{\text{cl}} \in \{0,1\}$ denote a question's total, knowledge-item, and clarity-item points under our checklist protocol (Section~\ref{sec:metrics}). \textbf{Overall Performance} (\%) is the mean of $s/10$; \textbf{Knowledge Score} (\%) is the mean of $s_{\text{kn}}/9$; \textbf{Clarity Score} (\%) is the mean of $s_{\text{cl}}$; and \textbf{Perfect Solve Rate} is the fraction with $s=10$.


\subsection{RQ1: Performance of Language Models} \label{sec:rq1}

We present results addressing the research questions above, starting with an end-to-end benchmarking of frontier models on RepoProbe. We report (1) aggregate performance over the full benchmark and (2) a single stratified summary that covers all three RepoProbe axes from Section~3---programming languages, code-understanding categories, and repository themes---using means and standard deviations across the 20 evaluated systems (Table~\ref{tab:table_RQ1_strata_summary}). This design preserves the benchmark's compositional structure without dedicating multiple pages to per-model breakdown tables.

\subsubsection{Aggregate Performance}

We summarize the aggregate performance of models using the four metrics defined in Section~\ref{sec:experiment_setup}. \Cref{tab:table_RQ1} shows a clear but non-uniform capability hierarchy.

{
\setlength{\abovecaptionskip}{1pt}
\setlength{\belowcaptionskip}{1pt}
\renewcommand{\arraystretch}{0.92}
\begin{table*}[t]
    \centering
    \caption{Language model performance on the four evaluation metrics, reported as percentages. For each metric, the \textbf{best-performing} model is marked in \textbf{bold} and the \uline{second-best} is \uline{underlined}.}
    \label{tab:table_RQ1}
    \resizebox{\textwidth}{!}{
    \begin{tabular}{llcccc}\toprule
        Category & Model & Overall Performance & Knowledge Score & Clarity Score & Perfect Solve Rate\\
        \midrule
        \multirow{13}{*}{Closed-weight models} & GPT-5.2 & \textbf{62.7} & \textbf{60.3} & \uline{84.7} & 26.0\\
         & GPT-5.4 & 60.4 & 57.6 & \textbf{85.6} & 24.2\\
         & Claude Opus 4.5 & 58.1 & 56.2 & 75.1 & 23.1\\
         & Claude Opus 4.6 & \uline{62.1} & \uline{60.2} & 79.0 & \textbf{27.5}\\
         & Claude Sonnet 4.6 & 60.2 & 58.3 & 78.2 & \uline{27.2}\\
         & Grok 4.2 & 56.8 & 54.8 & 75.0 & 23.4\\
         & Gemini 3 Flash & 55.4 & 53.6 & 71.6 & 21.8\\
         & Gemini 3 Pro & 51.3 & 49.3 & 71.2 & 17.6\\
         & Gemini 3.1 Pro & 54.1 & 52.2 & 70.9 & 21.2\\
         & Qwen3-Max & 27.3 & 26.1 & 39.3 & 5.6\\
         & Qwen3.5-Plus & 50.0 & 48.1 & 66.9 & 17.8\\
         & Doubao-Seed-1.8 & 43.5 & 42.4 & 55.1 & 12.8\\
         & Doubao-Seed-2.0 & 46.8 & 45.1 & 62.2 & 14.6\\
        \midrule
        \multirow{7}{*}{Open-weight models} & GLM-4.7 & 50.7 & 49.3 & 64.8 & 18.4\\
         & GLM-5 & 54.8 & 52.9 & 72.2 & 21.3\\
         & DeepSeek-V3.2 & 52.9 & 51.3 & 68.7 & 18.8\\
         & Kimi-K2 & 45.9 & 44.1 & 61.3 & 15.6\\
         & Kimi-K2.5 & 53.7 & 51.8 & 71.1 & 19.9\\
         & MiniMax-M2.1 & 45.0 & 43.6 & 57.8 & 12.8\\
         & MiniMax-M2.5 & 44.6 & 42.8 & 61.2 & 14.5\\
        \bottomrule
    \end{tabular}
    }
\end{table*}
}
 
Overall, frontier closed-weight models occupy most of the top positions, while open-weight systems remain competitive but generally lag in checklist coverage and perfect-solve frequency. More importantly, the two-time-point design reveals that progress across model generations is \textbf{often positive but not monotonic}: newer releases such as \texttt{Claude Opus 4.6}, \texttt{Gemini 3.1 Pro}, \texttt{GLM-5}, and \texttt{Kimi-K2.5} improve over their earlier counterparts, yet this trend is not universal, as shown by \texttt{GPT-5.4} and \texttt{MiniMax-M2.5}. This suggests that general frontier progress does not automatically translate into better repository-level understanding.

Across all models, three shared performance signatures emerge. First, \textbf{clarity consistently outpaces technical correctness}: models often produce well-structured explanations while missing required checklist items. Second, \textbf{perfect solves remain moderate} even for frontier systems, indicating that repository-level Q\&A is frequently limited by evidence grounding and cross-file dependency tracing rather than fluent narration. Third, the score distribution exhibits a \textbf{non-trivial low-score tail}, where a noticeable fraction of tasks receive near-zero credit rather than degrading gracefully.

\subsubsection{Stratified Performance: Languages, Categories, and Themes}

We assess cross-language robustness, structural difficulty across code-understanding categories, and domain sensitivity across repository themes using the compact summary in \Cref{tab:table_RQ1_strata_summary}. Each row reports the mean \textbf{Overall Performance} across all 20 models, with the standard deviation in parentheses; for the three categories we additionally report the mean \textbf{Perfect Solve Rate}, which is especially diagnostic for end-to-end completion.

{
\setlength{\abovecaptionskip}{1pt}
\setlength{\belowcaptionskip}{1pt}
\begin{table}[t]
\centering
\caption{Stratified RepoProbe performance summarized across all 20 evaluated models: mean Overall Performance (\%) with standard deviation in parentheses. Mean Perfect Solve Rate (\%) is also included for code-understanding categories. ``---'' indicates that the stratum only uses the overall column.}
\label{tab:table_RQ1_strata_summary}
\footnotesize
\setlength{\tabcolsep}{4pt}
\begin{tabular}{@{}llrr@{}}
\toprule
\textbf{Axis} & \textbf{Stratum} & \textbf{Overall (\%)} & \textbf{Perfect (\%)} \\
\midrule
\multirow{3}{*}{Category}
  & Project Architecture & $53.8$ $(7.9)$ & $19.2$ $(5.7)$ \\
  & Business Logic       & $52.3$ $(9.4)$ & $15.8$ $(5.4)$ \\
  & Implementation Det.\  & $51.2$ $(8.2)$ & $19.4$ $(5.6)$ \\
\midrule
\multirow{8}{*}{Theme}
  & LLM Framework   & $51.8$ $(8.0)$ & --- \\
  & Cloud DevOps    & $55.4$ $(7.6)$ & --- \\
  & Developer SDK   & $50.3$ $(8.6)$ & --- \\
  & Editor UI       & $50.7$ $(8.7)$ & --- \\
  & Monitoring Tools& $55.1$ $(8.6)$ & --- \\
  & Desktop Apps    & $47.3$ $(9.0)$ & --- \\
  & Data Processing & $47.6$ $(7.6)$ & --- \\
  & Security        & $56.0$ $(8.9)$ & --- \\
\midrule
\multirow{15}{*}{Language}
  & Python       & $53.1$ $(7.8)$ & --- \\
  & TypeScript   & $57.6$ $(8.7)$ & --- \\
  & Lua          & $55.6$ $(9.5)$ & --- \\
  & Shell        & $52.2$ $(7.7)$ & --- \\
  & Go           & $52.1$ $(7.8)$ & --- \\
  & Rust         & $48.2$ $(9.7)$ & --- \\
  & QML          & $38.4$ $(8.8)$ & --- \\
  & C\#          & $47.6$ $(7.5)$ & --- \\
  & Swift        & $43.3$ $(9.2)$ & --- \\
  & C++          & $47.4$ $(11.4)$ & --- \\
  & Ruby         & $48.8$ $(6.7)$ & --- \\
  & JavaScript   & $46.5$ $(8.9)$ & --- \\
  & Java         & $56.4$ $(9.7)$ & --- \\
  & C            & $51.9$ $(6.8)$ & --- \\
  & PHP          & $45.4$ $(8.3)$ & --- \\
\bottomrule
\end{tabular}
\vspace{-4pt}
\end{table}
}

\textbf{Programming languages.} Mean scores span a wide band (e.g., QML at $38.4\%$ vs.\ TypeScript at $57.6\%$), with several ecosystems, notably QML, Swift, and PHP falling well below the benchmark-wide average. High cross-model standard deviations on languages such as C++ also indicate uneven capability across systems. Together with the aggregate results in \Cref{tab:table_RQ1}, this pattern suggests transferable strengths for mainstream stacks but persistent weakness in long-tail languages.

\textbf{Code-understanding categories.} Mean overall scores are similar across categories, but \textbf{perfect solves} are lowest for \textit{Business Logic}, indicating that fully satisfying all checklist items is hardest when tasks require causal and cross-component reasoning even when partial credit remains common elsewhere.

\textbf{Repository themes.} Cross-model means separate ``easier'' domains (e.g., \textit{Security} and \textit{Cloud DevOps}) from more challenging ones (e.g., \textit{Desktop Apps} and \textit{Data Processing}), with the remaining themes in between. This aligns with heterogeneous real-world architectural conventions while still showing that no single theme explains overall leaderboards: strong models remain competitive across themes relative to weaker ones, consistent with \Cref{tab:table_RQ1}.

\begin{tcolorbox}[colback=gray!5,colframe=gray!60,toprule=0.5pt,bottomrule=0.5pt,top=2pt,bottom=2pt,title=Answer to RQ1]
RepoProbe reveals a clear but non-monotonic frontier capability hierarchy. Newer SOTA releases often outperform earlier ones, but the gains are inconsistent across model families, indicating that general model progress does not automatically translate into stronger repository-level understanding. Moreover, even the best-performing system reaches only the low-60\% range, leaving substantial room for improvement.
\end{tcolorbox}

\subsection{RQ2: Stability and Interpretability of Evaluation Metrics} 
\label{sec:rq2}

To assess the stability and reproducibility of our evaluation framework, we conducted a rigorous comparative analysis between the proposed \textbf{Checklist-Based Verification Protocol} and the traditional LLM-as-a-Judge scalar scoring method. Our analysis reveals that while scalar scoring is widely used, it suffers from critical instability and a lack of interpretability when applied to complex repository-level Q\&A tasks.

\subsubsection{Macro-level Analysis: The Illusion of Stability}

At an aggregated level, scalar scoring based on a 5-dimensional Likert scale appears moderately stable. However, a deeper statistical analysis exposes significant volatility that is masked by average scores. As shown in \Cref{tab:macro_variance_compare}, we compared the performance of three SOTA models using both metrics across 5 independent evaluation runs.
 
The scalar baseline follows SWE-QA \citep{peng2025sweqa} LLM-as-a-Judge prompt: given the question, the reference answer, and the candidate answer, the judge assigns holistic scores on a 5-dimensional Likert rubric. For both checklist-based verification and scalar scoring in this experiment, we use \texttt{Claude Sonnet 4.5} as the fixed evaluator under the same generation settings and retry policy, ensuring consistent evaluation across all runs.

{
\setlength{\textfloatsep}{8pt plus 2pt minus 2pt}
\setlength{\dbltextfloatsep}{8pt plus 2pt minus 2pt}
\setlength{\abovecaptionskip}{1pt}
\setlength{\belowcaptionskip}{1pt}
\begin{table}[t]
    \centering
    \caption{\textbf{Reliability Comparison: Scalar vs. Checklist.} Across three models, Checklist-Based Verification consistently demonstrates superior stability, with standard deviations roughly half that of Scalar Scoring.}
    \label{tab:macro_variance_compare}
    \footnotesize
    \renewcommand{\arraystretch}{0.95}
    \setlength{\tabcolsep}{3.2pt}
    \resizebox{\columnwidth}{!}{%
    \begin{tabular}{l|ccc|ccc}
    \toprule
    & \multicolumn{3}{c|}{\textbf{Scalar Scoring (5-Dimensions)}} & \multicolumn{3}{c}{\textbf{Checklist-Based Verification}} \\
    \midrule
    \textbf{Model} & \textbf{Score Rate} & \textbf{StdDev} & \textbf{Range} & \textbf{Score Rate} & \textbf{StdDev} & \textbf{Range} \\
    \midrule
    GPT-5.2 & 91.3\% & 2.9\% & 6.9\% & 76.3\% & 1.7\% & 3.8\% \\
    Claude Opus 4.5 & 85.4\% & 3.5\% & 8.3\% & 71.8\% & 1.5\% & 3.2\% \\
    Gemini 3 Pro & 82.2\% & 3.3\% & 7.8\% & 65.6\% & 2.4\% & 5.2\% \\
    \bottomrule
    \end{tabular}%
    }
    \vspace{-4pt}
\end{table}
}

The results demonstrate a clear stability gap. The \textbf{Checklist-Based Verification} method consistently exhibits a lower Standard Deviation and a narrower Range compared to Scalar Scoring. Most notably, the scalar metric's range reaches up to 8.3\% for \texttt{Claude Opus 4.5}. This implies that under the traditional Likert scale, a model's evaluation could fluctuate by nearly a full letter grade purely due to stochastic noise in the judge's interpretation, rendering it unsuitable for precise benchmarking. In contrast, the rubric-guided nature of the checklist forces the evaluator to converge on evidence-backed technical checks, cutting this measurement error by more than half.

{
\begin{table*}[t]
    \centering
    \caption{\textbf{Instability of Scalar Scoring (Micro-Analysis).} 
    \textbf{Case 1 (Variance):} Independent judges fundamentally disagree on whether a claim is an error or a feature, causing massive score fluctuation (40\%), while checklist-based verification provides consistent identification of specific missing components. 
    \textbf{Case 2 (Uninterpretability):} Even with identical scores, judges provide contradictory directional feedback (one requesting more detail, one requesting less), making the metric unactionable for optimization, whereas checklist-based verification delivers actionable feedback by pinpointing concrete deficiencies.}
    \label{tab:variance_cases}
    \small
    \setlength{\tabcolsep}{4pt}
    \begin{tabularx}{\textwidth}{l l X X X}
    \toprule
    \textbf{Case Type} & \textbf{Model} & \textbf{Scalar Run A} & \textbf{Scalar Run B} & \textbf{Checklist} \\
    \midrule
    \textbf{Case 1} & \texttt{Claude Opus 4.5} & \textbf{Score: 50\%} & \textbf{Score: 90\%} & \textbf{Score: 70\%} \\
    \textit{High Variance} & (\texttt{Pulse}) & \textit{``Critical correctness issue: claims fully supports... while reference says possible solution.''} & \textit{``Highly detailed and technically accurate... correctly explains why...''} & \textit{``Provides extensive detail about Solution 1... However, it COMPLETELY MISSES Solution 2.'} \\
    (Scalar Diff: 40\%) & & \textcolor{red}{$\rightarrow$ Interpretation: Contradiction} & \textcolor{green!60!black}{$\rightarrow$ Interpretation: Elaboration} & \textcolor{blue}{$\rightarrow$ Consistent: Missing Component} \\
    \midrule
    \textbf{Case 2} & \texttt{GPT-5.2} & \textbf{Score: 80\%} & \textbf{Score: 80\%} & \textbf{Score: 30\%} \\
    \textit{Uninterpretability} & (\texttt{adk-python}) & \textit{``Lacks the specific data structure example shown in the reference.''} & \textit{``More verbose and complex than the reference... added complexity.''} & \textit{``Incomplete file path; missing storage mechanism explanation.''} \\
    (Scalar Diff: 0\%) & & \textcolor{orange}{$\rightarrow$ Feedback: Add Detail (+)} & \textcolor{orange}{$\rightarrow$ Feedback: Reduce Detail (-)} & \textcolor{blue}{$\rightarrow$ Consistent: Specific Gaps} \\
    \bottomrule
    \end{tabularx}
\end{table*}
}

\subsubsection{Micro-level Case Studies: Anatomy of Variance}

To understand the qualitative drivers of this instability, we analyzed specific instances where scalar scoring failed to provide consistent signals. We identified two failure modes. \textit{Interpretation Drift} leads to high variance, while \textit{Directional Inconsistency} leads to uninterpretability.

\textbf{Failure Mode 1: Interpretation Drift: High Variance.} Scalar judges often struggle to consistently evaluate answers that deviate from the strict scope of the reference. As illustrated in \Cref{tab:variance_cases}, Case 1 evaluates \texttt{Claude Opus 4.5} on the \texttt{Pulse} repository, where independent runs produced conflicting interpretations of the same claim. One judge viewed the model's assertion about QEMU Guest Agent support as a contradiction and assigned 50\%, while another viewed it as a valid elaboration and assigned 90\%. This near-doubling of the score highlights the scalar metric's inability to objectively distinguish between \textit{hallucination} and \textit{insight}. By contrast, our checklist-based approach eliminates this ambiguity through criterion decomposition: across five independent evaluation runs, all judges consistently identified the same specific deficiency, namely missing the second solution component, and converged on identical reasoning without interpretation variance. This stability stems from evaluating \textit{verifiable facts} rather than \textit{subjective quality}.

\textbf{Failure Mode 2: Directional Inconsistency: Uninterpretability.}
Even when scalar scores converge, they may provide contradictory feedback, making them useless for model improvement. In \Cref{tab:variance_cases}, Case 2 evaluates \texttt{GPT-5.2} on \texttt{adk-python}. Two judges assigned the same score of 80\%, yet they justified it in mutually exclusive ways. One judge argued that the answer lacked specific examples and therefore needed \textit{more} detail, while the other argued that the answer was more verbose than the reference and therefore needed \textit{less} detail. A developer relying on this metric would receive conflicting gradients for optimization. The checklist approach resolves this uninterpretability by providing \textit{granular, reproducible feedback}: all five evaluation runs consistently identified the same three specific gaps. This enables developers to address concrete deficiencies rather than reconciling contradictory holistic judgments.

\textbf{The Checklist Advantage.}
RepoProbe mitigates these variances by replacing subjective \textit{ratings} with objective \textit{verifications}. Instead of asking ``Is the answer correct?'' in a scalar form, our protocol asks specific binary questions, such as \textit{``Does the answer include a code example for InMemoryCredentialService?''} This constraint eliminates the ambiguity observed in Case 2 and ensures that evaluation feedback is consistent and actionable.

\begin{tcolorbox}[colback=gray!5,colframe=gray!60,toprule=0.5pt,bottomrule=0.5pt,top=2pt,bottom=2pt,title=Answer to RQ2]
Our checklist-based verification protocol delivers more stable and reproducible evaluation for repository-level Q\&A task by decomposing each task into evidence-backed technical checks, reducing measurement noise and improving interpretability. In contrast, traditional LLM-as-a-Judge scalar scoring can exhibit interpretation drift and directional inconsistency, resulting in unstable scores and contradictory feedback.
\end{tcolorbox}

\subsection{RQ3: Architectural Insight over Edit Bias} \label{sec:rq3}
This research question investigates the qualitative advantages of discussion-based tasks over traditional code generation. We revisit the core \textit{Edit Bias} failure mode to demonstrate why architectural understanding is a prerequisite for safe code generation.

In our qualitative analysis of model outputs, we frequently observed that models performing well on traditional pass/fail code generation metrics often failed RepoProbe's architectural queries. We categorize this as \textit{Edit Bias}, where models attempt to solve architectural questions by generating new code or configurations rather than investigating the existing system design.

\begin{table}[t]
\centering
\caption{Breakdown of Failure Modes across Top Models}
\captionsetup{skip=5pt}
\label{tab:failure_taxonomy}
\fontsize{7.8}{9.2}\selectfont
\setlength{\tabcolsep}{4pt}
\begin{tabular*}{\columnwidth}{@{\extracolsep{\fill}}>{\raggedright\arraybackslash}p{3.35cm}cccc@{}}
\toprule
\textbf{Failure Mode} & \textbf{GPT-5.2} & \textbf{Claude} & \textbf{Gemini} & \textbf{Avg.} \\
\midrule
\textbf{Edit Bias} & \textbf{10.4\%} & \textbf{13.0\%} & \textbf{24.0\%} & \textbf{15.8\%} \\
\multicolumn{1}{>{\raggedright\arraybackslash}p{3.35cm}}{\textit{(Focuses on code edits over understanding)}} & & & & \\
\addlinespace[1pt]
\textbf{Shallow Explanation} & 38.5\% & 21.0\% & 25.0\% & 28.2\% \\
\multicolumn{1}{>{\raggedright\arraybackslash}p{3.35cm}}{\textit{(Vague or incomplete technical details)}} & & & & \\
\addlinespace[1pt]
\textbf{Misinterpretation} & 29.2\% & 43.0\% & 26.0\% & 32.7\% \\
\multicolumn{1}{>{\raggedright\arraybackslash}p{3.35cm}}{\textit{(Finds code but analyzes incorrectly)}} & & & & \\
\addlinespace[1pt]
\textbf{Context Miss} & 17.7\% & 21.0\% & 18.0\% & 18.9\% \\
\multicolumn{1}{>{\raggedright\arraybackslash}p{3.35cm}}{\textit{(Fails to retrieve relevant files)}} & & & & \\
\addlinespace[1pt]
\textbf{Hallucination} & 3.1\% & 2.0\% & 6.0\% & 3.7\% \\
\multicolumn{1}{>{\raggedright\arraybackslash}p{3.35cm}}{\textit{(References non-existent entities)}} & & & & \\
\bottomrule
\end{tabular*}
\vspace{-4pt}
\end{table}

\begin{figure*}[t]
\centering
\small
\begin{tabular}{p{0.48\textwidth} p{0.48\textwidth}}
\toprule
\multicolumn{2}{p{0.96\textwidth}}{\textbf{User Query:} \textit{``Is it possible to add a new config for allowing the user to specify the log saving location?''}} \\
\midrule
\multicolumn{1}{c}{\textbf{\textcolor{red!70!black}{Instruction Following (Shallow)}}} & \multicolumn{1}{c}{\textbf{\textcolor{green!40!black}{Architectural Insight (Deep)}}} \\
\midrule
\textbf{Model Response (GPT-5.2)} & \textbf{Reference Answer (Maintainer)} \\
\textit{``Yes! Here is how to add a `log\_dir` config...''} & \textit{``The problem is not the path, it's the symlink privilege.''} \\
\vspace{0.1em}
\textbullet\ Generates CLI flag code & \textbullet\ Ignores request for config \\
\textbullet\ Modifies \texttt{cli\_tools\_click.py} & \textbullet\ Replaces \texttt{os.symlink} with \texttt{shutil.copy2} \\
\vspace{0.1em}
\textbf{Outcome:} User changes path, but \textbf{still crashes} because symlink creation is hardcoded. & \textbf{Outcome:} Log system works without admin rights. \\
\midrule
\textbf{Checklist Evaluation (0/4)} & \textbf{Checklist Evaluation (4/4)} \\
\textcolor{red}{\ding{55}} Identifies root cause (Symlink Privilege) & \textcolor{green!60!black}{\ding{51}} Identifies root cause (Symlink Privilege) \\
\textcolor{red}{\ding{55}} Solution addresses root cause & \textcolor{green!60!black}{\ding{51}} Solution addresses root cause \\
\textcolor{red}{\ding{55}} Replaces \texttt{os.symlink} with copy & \textcolor{green!60!black}{\ding{51}} Replaces \texttt{os.symlink} with copy \\
\bottomrule
\end{tabular}
\caption{\textbf{Case Study: The ``Leading Question'' Trap.} RepoProbe uses Discussion data to expose models that blindly follow user instructions (Left) versus those that understand system architecture (Right). Traditional benchmarks might reward Path A for ``correct code generation,'' while the Checklist Verification correctly penalizes it for failing to solve the actual problem.}
\label{fig:rq3_case_study}
\end{figure*}

To quantify this, we employed an LLM-as-a-Judge using \texttt{Gemini 3 Flash} to categorize all failure cases with a score below 60\% across 500 questions into five distinct failure modes. These five categories were defined by the authors from recurring patterns observed during qualitative inspection of low-scoring answers. We manually inspected the cases assigned to \textit{Edit Bias} to verify that they reflected premature code-generation behavior rather than evidence-grounded architectural analysis. The results, summarized in \Cref{tab:failure_taxonomy}, reveal a significant trend.
While \textit{Shallow Explanation} and \textit{Misinterpretation} remain common across all models, \textit{Edit Bias} emerges as a notable failure pattern in SOTA models. As shown in the table, this specific failure mode accounts for a significant portion of all failure cases, ranging from \textbf{10.4\%} in \texttt{GPT-5.2} to \textbf{24.0\%} in \texttt{Gemini 3 Pro}. This indicates that even highly capable models, when faced with open-ended repository inquiries, often default to \textit
{Edit Bias}, prioritizing immediate code generation over the necessary structural analysis. A representative example involves the \texttt{adk-python} repository, as illustrated in \Cref{fig:rq3_case_study}.

\textbf{The Scenario:} A user asks, ``Is it possible to add a new config for allowing the user specify the log saving location?'' because they are encountering a Windows privilege error \texttt{WinError 1314} when the system tries to create a symlink.

\textbf{The Model Failure (GPT-5.2):} The model correctly identifies that the code \textit{can} be modified to add a config. It generates a long, plausible implementation plan involving modifying the file \texttt{cli\_tools\_click.py} and adding a CLI flag. However, it fails to understand the \textit{why} behind the user's request: the user is only asking for a config because they want to bypass a symlink error. The model's proposed solution, adding a config, is an instance of \textit{Edit Bias} that adds engineering cost without solving the root cause, namely the privilege requirement for symlinks on Windows.
 
\textbf{The Reference Solution (Architectural Insight):} The correct approach, derived from the repository maintainer's answer, is not to add a new feature, but to simply change the file operation from \texttt{os.symlink} to \texttt{shutil.copy2}. This requires understanding that the \textit{intent} of the question is to fix the crash rather than satisfy the \textit{literal request} to add a config.

\textbf{Implication:} Discussion-based benchmarks like RepoProbe penalize models that blindly follow instructions without understanding the system's constraints. A checklist-based verification captures this nuance by verifying \textit{``Does the solution address the symlink privilege root cause?''} as a Yes/No check. Models that merely generate the requested config code fail this check, whereas they might pass a unit test that only checks whether the config flag exists. This confirms that RepoProbe effectively distinguishes between coding capability and engineering judgment.

\begin{tcolorbox}[colback=gray!5,colframe=gray!60,toprule=0.5pt,bottomrule=0.5pt,top=2pt,bottom=2pt,title=Answer to RQ3]
Discussion-based tasks offer a distinct advantage by exposing the \textit{Edit Bias} failure mode: models often over-engineer solutions because they lack the architectural context to identify simple, low-code fixes. RepoProbe forces models to align with the system's design philosophy rather than just satisfying a functional unit test, providing a better proxy for the safety and robustness of autonomous agents.
\end{tcolorbox}
\section{Threats to Validity}
Threat in \textbf{data source}: Compared with GitHub Issues, GitHub Discussions contain more open-ended and community-facing content, which raises the risk of including threads that are only weakly related to repository understanding. We control this threat by restricting data collection to popular repositories with active Discussions usage, requiring answered threads, and applying multiple filtering stages that remove non-code discussions and retain only samples with clear code-understanding intent. We further verify answerability against repository code and perform expert inspection on a validated subset to ensure that the final benchmark questions are genuinely grounded in repository understanding rather than incidental community chatter. Threat in \textbf{benchmark coverage}: performance conclusions may depend on the programming languages, repository themes, and projects included in the benchmark. We mitigate this threat by constructing RepoProbe benchmark from 50 repositories spanning 15 programming languages and multiple repository themes, and by reporting stratified analyses across programming languages, code-understanding categories, and repository themes. This design reduces the risk that our conclusions are driven by a single ecosystem or task subtype and better supports the evaluation across heterogeneous software settings.

\section{Conclusion}
This work introduces \textbf{RepoProbe}, a Q\&A benchmark for repository-level code understanding grounded in real GitHub Discussions, together with a \textbf{Checklist-Based Verification Protocol} for evaluating open-ended repository answers. Across frontier LLMs, RepoProbe reveals a persistent gap between fluent explanation and evidence-grounded correctness, while our reliability study shows that checklist-based verification is more stable and actionable than scalar judging. The benchmark also surfaces \textit{Edit Bias} as a recurring failure mode in open-ended repository understanding. Future work can extend RepoProbe toward richer evidence grounding, such as multimodal artifacts and explicit file-span citations.
\section*{Acknowledgement}

This work was supported in part by the Zhejiang Pioneer (Jianbing) Project (2026C01024) and in part by Ningbo Global Innovation Center, Zhejiang University (Grant No. NBRY2025X0018).
\section*{Data Availability Statement}
All code and dataset that led to the results in the paper are publicly available at \cite{paperdataset}.

\bibliographystyle{ACM-Reference-Format}
\bibliography{paper}

\end{document}